\documentclass[12pt]{article}
\usepackage{scicite}
\usepackage{times}
\usepackage{amsmath}
\newenvironment{sciabstract}{%
\begin{quote} \bf}
{\end{quote}}

\usepackage{graphicx}
\usepackage{subcaption}
\usepackage{hyperref}

\def\TXS/{TXS~0506+056}  
\def\ICA/{IceCube-170922A}
\def\FermiLAT/{{\it Fermi}-LAT}

\title{Neutrino emission from the direction of the blazar TXS~0506+056 prior to the IceCube-170922A alert} 

\author
{IceCube Collaboration\footnote{The full list of collaboration members and their affiliations is included in the supplementary material}
\footnote{Correspondence to analysis@icecube.wisc.edu}\\
}

\date{}

\begin{document} 
\baselineskip15pt

\maketitle 

\begin{sciabstract}
A high-energy neutrino event detected by IceCube on 22 September 2017 was coincident
in direction and time with a gamma-ray flare from the blazar TXS 0506+056. Prompted by this association, we investigated 9.5 years of IceCube neutrino observations to search for excess emission at the position of the blazar. We found an excess of high-energy neutrino events with respect
to atmospheric backgrounds at that position between September 2014 and March 2015. Allowing for time-variable flux, this constitutes ${3.5 \sigma}$ evidence for neutrino emission from the direction of \TXS/, independent of and prior to the 2017 flaring episode. This suggests that blazars are the first identifiable sources of the high-energy astrophysical neutrino flux.
\end{sciabstract}

The origin of the highest-energy cosmic rays is believed to be extragalactic \cite{Aab:2017tyv}, 
but their acceleration sites remain unidentified.
High-energy neutrinos are expected to be produced in or near the acceleration sites when cosmic rays interact with matter and ambient light, producing charged mesons that decay into neutrinos and other particles.
Unlike cosmic rays, neutrinos can travel through the Universe unimpeded by interactions with other particles and undeflected by magnetic fields, providing a means to identify and study the extreme environments producing cosmic rays \cite{Reines:1960we}.
Blazars, a class of active galactic nuclei with powerful relativistic jets pointed close to our line of sight  \cite{1995PASP..107..803U}, 
are prominent candidate sources of such high-energy neutrino emission
\cite{1995APh.....3..295M,1997ApJ...488..669H,2003APh....18..593M,Murase2017,Petro2016,Guepin2017}. The electromagnetic emission of blazars is observed to be highly variable on time-scales from minutes to years \cite{1997ARA&A..35..445U}. 

The IceCube Neutrino Observatory \cite{2017JInst..12P3012A} 
is a high-energy neutrino detector
occupying an instrumented volume of 1~km$^{3}$ within the Antarctic ice sheet at the Amundsen-Scott South Pole Station. The detector consists of an array of 86 vertical strings,
nominally spaced 125~m apart and 
descending to a depth of approximately 2450~m in the ice. The bottom 1~km of each string is equipped with 60 optical sensors that record Cherenkov light emitted by relativistic charged particles passing through the optically transparent ice. When high-energy muon neutrinos interact with the ice, they can create relativistic muons that travel many kilometers, 
creating a track-like series of Cherenkov photons recorded when they pass through the array.
This allows the reconstruction of the original neutrino direction with a median angular uncertainty of $0.5^{\circ}$ for a neutrino energy of $\sim 30$~TeV (or  $0.3^{\circ}$ at 1~PeV) 
\cite{2017ApJ...835..151A,Aartsen:2013zka}. 

IceCube discovered the existence of a diffuse flux of high-energy astrophysical neutrinos in 2013
\cite{2013Sci...342E...1I,2014PhRvL.113j1101A}. 
Measurements of the energy spectrum have since been refined
\cite{2015ApJ...809...98A,2016ApJ...833....3A }, 
indicating that the neutrino spectrum extends above several PeV. 
However, analyses of neutrino observations have not succeeded in identifying individual sources of high-energy neutrinos \cite{2017ApJ...835..151A,2015ApJ...807...46A}. 
This suggests that the sources are distributed
 across the sky and that even the brightest individual sources contribute only a small fraction of the total observed flux. 

Recently, the detection of a high-energy neutrino by IceCube, together with observations in gamma rays and at other wavelengths, 
indicates that a blazar, \TXS/,
located at right ascension (RA) $77.3582^{\circ}$ and declination (Dec) $+5.69314^{\circ}$ (J2000 equinox) \cite{2015Ap&SS.357...75M} 
may be an individually identifiable source of high-energy neutrinos\cite{MMpaper}. 
The neutrino-candidate event, \ICA/, was detected on 22 September 2017,
selected by the Extremely High Energy (EHE) online event filter
\cite{Aartsen:2016lmt}, 
and reported as a public alert
\cite{GCN21916}.
EHE alerts are currently sent at a rate of about four per year, and are based on well-reconstructed, high-energy muon-track events. The selection threshold is set so that approximately half of the events are estimated to be astrophysical neutrinos, the rest being atmospheric background events. After the alert was sent, further studies refined the directional reconstruction, with best-fitting coordinates of RA 77.43$_{-0.65}^{+0.95}$
and Dec $+5.72_{-0.30}^{+0.50}$ (deg, J2000, 90\% containment region). The most probable neutrino energy was estimated to be 290 TeV, with a 90\% confidence level lower limit of 183 TeV \cite{MMpaper}.

It was soon determined that the direction of \ICA/ was consistent with the location of \TXS/ and coincident with a state of enhanced gamma-ray activity observed since April 2017 \cite{ATEL10791}
by the Large Area Telescope (LAT) on the {\it Fermi Gamma-ray Space Telescope}\cite{2009ApJ...697.1071A}.
Follow-up observations of the blazar led to the detection of gamma rays with energies up to 400~GeV by the Major Atmospheric Gamma Imaging Cherenkov (MAGIC) Telescopes 
\cite{2012APh....35..435A,ATEL10817}.  \ICA/ and the 
electromagnetic observations are described in detail in \cite{MMpaper}.  The significance of the spatial and temporal coincidence of the high-energy neutrino and the blazar flare is estimated to be at the $3\sigma$ level \cite{MMpaper}.
On the basis of this result, we consider the hypothesis that the blazar \TXS/ has been a source of high-energy neutrinos beyond that single event.

\section*{Searching for neutrino emission}

IceCube monitors the whole sky and has maintained essentially continuous observations since 5 April 2008.
Searches for neutrino point sources using two model-independent methods, a time-integrated and a time-dependent unbinned maximum likelihood analysis, have previously been published
for the data collected between 2008 and 2015
\cite{2017ApJ...835..151A,2012ApJ...744....1A,2015ApJ...807...46A}.
Here, we analyze the same 7-year data sample supplemented with additional data collected from May 2015 until October 2017 \cite{Aartsen:2016lmt}. The data span 9.5 years and consist of six distinct periods, corresponding to changing detector configurations, data-taking conditions, and improved event selections (Table~\ref{tab:data_samples}).

\begin{table}[ht]
\begin{center}
\begin{tabular}{l c c}
\hline
\hline
Sample & Start & End \\
\hline
IC40  & 2008 Apr\,\,\,\,\,5  & 2009 May 20 \\
IC59  & 2009 May 20 & 2010 May 31 \\
IC79  & 2010 May 31 & 2011 May 13 \\
IC86a & 2011 May 13 & 2012 May 16 \\
IC86b & 2012 May 16 & 2015 May 18 \\
IC86c & 2015 May 18 & 2017\, Oct\, 31 \\
\hline
\end{tabular}
\caption{{\bf IceCube neutrino data samples.} Six data-taking periods make up the full 9.5-year data sample. Sample numbers correspond to the number of detector strings that were operational. During the first three periods, the detector was still under construction. The last three periods correspond to different data-taking conditions and/or event selections with the full 86-string detector.}
\label{tab:data_samples}
\end{center}
\end{table}

The northern sky, where \TXS/ is located, 
is observed through Earth
by IceCube. Approximately 70,000 neutrino-induced muon tracks are recorded each year from this hemisphere of the sky after passing the final event selection criteria. Fewer than 1\% of these events originate from astrophysical neutrinos;
the vast majority are background events caused by neutrinos of median energy $\sim 1$~TeV 
created in cosmic-ray interactions in the atmosphere over other locations on Earth.
However, for an astrophysical muon-neutrino flux where the differential number of neutrinos with energy $E$ scales like $dN/dE \sim E^{-2}$,
the distribution of muon energies is different than for the background atmospheric neutrino flux, which scales as $\sim E^{-3.7}$ \cite{2016ApJ...833....3A }. 
This allows for further discriminating power in point source searches besides directional-only excesses.

A high-significance point-source detection \cite{2017ApJ...835..151A,2015ApJ...807...46A} 
can require between as few as two or three, or as many as 30,
signal events to stand out from the background, depending on 
the energy spectrum and the clustering of events in time.
To search for a neutrino signal at the coordinates of \TXS/, we apply the standard time-integrated analysis \cite{2008APh....29..299B} 
and time-dependent analysis \cite{timeDepMethod2010APh....33..175B} 
that have been used in past searches \cite{2017ApJ...835..151A,2012ApJ...744....1A,2015ApJ...807...46A}. 
The time-integrated analysis 
uses an unbinned maximum-likelihood ratio method to search for an excess number of events 
consistent with a point source at a specified location, given the angular distance and angular uncertainty of each event. Energy information is included in the definition of the likelihood, assuming a power-law energy spectrum, $E^{-\gamma}$,
with the spectral index $\gamma$ as a fitted parameter.
The model parameters are correlated and are expressed as a pair, ($\Phi_{100}$, $\gamma$), where $\Phi_{100}$ is the flux normalization at 100~TeV.  The time-dependent analysis 
uses the same formulation of the likelihood but searches for clustering in time as well as space by introducing an additional  
time profile. It is performed separately for two different generic profile shapes: a Gaussian-shaped time window and a box-shaped time window. Each analysis varies the central time of the window, $T_0$, and the duration $T_\mathrm{W}$ (from seconds to years) of the potential signal to find the four parameters $(\Phi_{100},\gamma,T_0,T_\mathrm{W})$ that maximize the likelihood ratio, which is defined as the test statistic $TS$. (For the Gaussian time window, $T_\mathrm{W}$ represents twice the standard deviation.) The test statistic includes a factor that corrects for the look-elsewhere effect arising from all of the possible time windows that could be chosen \cite{Supplementary}.

For each analysis method (time-integrated and time-dependent), a robust significance estimate is obtained by performing the identical analysis on trials with randomized data sets. These are produced by randomizing the event times and re-calculating the RA coordinates within each data-taking period.  The resultant $P$ value is defined as the fraction of randomized trials yielding a value of $TS$ greater than or equal to the one obtained for the actual data.

Because the detector configuration and event selections changed as shown in Table~\ref{tab:data_samples}, the time-dependent analysis is performed by operating on each data-taking period separately. (A flare that spans a boundary between two periods could be partially detected in either period, but with reduced significance.)
An additional look-elsewhere correction then needs to be applied for a result in an individual data segment, given by the ratio of the total 9.5~year observation time to the observation time of that data segment \cite{Supplementary}.

\section*{Neutrinos from the direction of \TXS/}

The results of the time-dependent analysis performed at the coordinates of
\TXS/ are shown in Fig.~1
for each of the six data
periods. One of the data periods, IC86b from 2012 to 2015, contains a significant excess
which is identified by both time-window shapes.  The excess consists of $13\pm5$ events above the expectation from the atmospheric background. The significance depends on the energies of the events, their proximity to the coordinates of \TXS/, and their clustering in time.  This is illustrated in Fig.~2,
which shows the time-independent weight of individual events in the likelihood analysis during the IC86b data period.

\begin{figure}[ht!]
\begin{center}
\includegraphics[width=\textwidth]{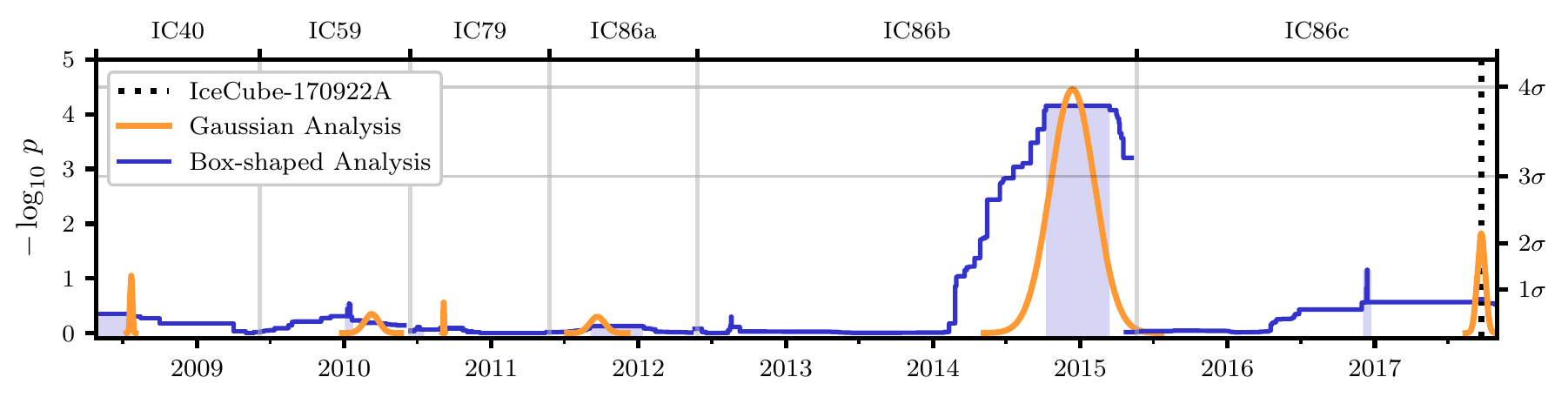}
\caption{{\bf Time-dependent analysis results.} The orange curve corresponds to the analysis using the Gaussian-shaped time profile. The central time $T_0$ and width $T_\mathrm{W}$ are plotted for the most significant excess found in each period, with the $P$ value of that result indicated by the height of the peak.  The blue curve corresponds to the analysis using the box-shaped time profile. The curve traces the outer edge of the superposition of the best-fitting time windows (durations $T_\mathrm{W}$) over all times $T_0$, with the height indicating the significance of that window.  In each period, the most significant time window forms a plateau, shaded in blue. The large blue band centered near 2015 represents the best-fitting 158-day time window found using the box-shaped time profile. The vertical dotted line in IC86c indicates the time of the \ICA/ event.}
\label{fig:time_dep_timeline}
\end{center}
\end{figure}

\begin{figure}[ht!]
\begin{center}
\includegraphics[width=\textwidth]{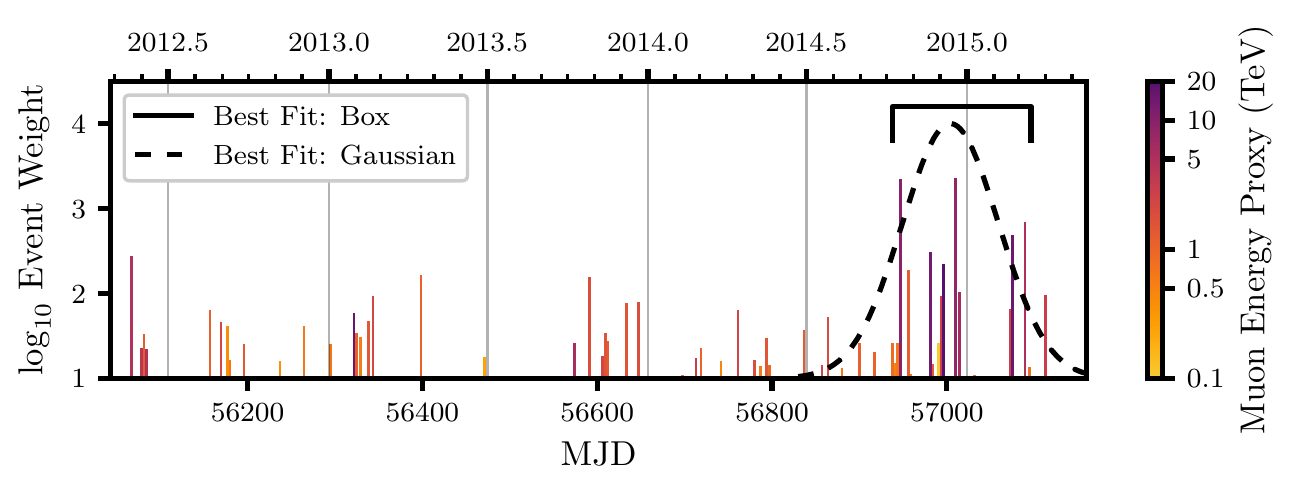}
\caption{{\bf Time-independent weight of individual events during the IC86b period.} Each vertical line represents an event observed at the time indicated by calendar year (top) or MJD (bottom). Overlapping lines are shifted by 1 to 2 days for visibility. The height of each line indicates the Event Weight: the product of the event's spatial term and energy term in the unbinned likelihood analysis evaluated at the location of \TXS/ and assuming the best-fitting spectral index $\gamma=2.1$ \cite{Supplementary}. The color for each event indicates an approximate value in units of TeV of the reconstructed muon energy (Muon Energy Proxy), which the analysis compares with expected muon energy distributions under different hypotheses. [A distribution for the true neutrino energy of a single event can also be inferred from the event's muon energy, see \cite{Supplementary}]. The dashed curve and the solid bracket indicate the best-fitting Gaussian and box-shaped time windows, respectively. The distribution of event weights and times outside of the best-fitting time windows is compatible with background.}
\label{fig:IC86b_event_timeline}
\end{center}
\end{figure}

The Gaussian time window is centered at 13 December 2014 [modified Julian day (MJD) 57004] with an uncertainty of $\pm 21$~days and a duration $T_\mathrm{W}=110_{-24}^{+35}$~days.  The best-fitting parameters
for the fluence $J_{100}=\int \Phi_{100}(t)dt$ and the 
spectral index are given by 
$E^2 J_{100} = (2.1^{+0.9}_{-0.7})\times10^{-4}$ TeV cm$^{-2}$
at 100~TeV and $\gamma=2.1\pm 0.2$, respectively.
The joint uncertainty on these parameters is shown in Fig.~3
along with a skymap showing the result of the time-dependent analysis performed at the location of \TXS/ and in its vicinity during the IC86b data period.

The box-shaped time window is centered 13 days later with duration $T_\mathrm{W}=158$~days (from MJD 56937.81 to MJD 57096.21, inclusive of contributing events at boundary times). For the box-shaped time window the uncertainties are discontinuous and not well-defined, but the uncertainties for the Gaussian window show that it is consistent with the box-shaped time window fit. Despite the different window shapes, which lead to different weightings of the events as a function of time, both windows identify the same time interval as significant.
For the box-shaped time window, the best-fitting parameters are similar to those of the Gaussian window, 
with fluence at 100~TeV and spectral index given by $E^2 J_{100} = (2.2^{+1.0}_{-0.8})\times10^{-4}$ TeV cm$^{-2}$ and $\gamma=2.2\pm 0.2$.
This fluence corresponds to an average flux over 158 days of $\Phi_{100} = (1.6^{+0.7}_{-0.6})\times10^{-15}$ TeV$^{-1}$ cm$^{-2}$ s$^{-1}$.

\begin{figure}[ht!]
\begin{center}
\begin{subfigure}[t]{0.48\textwidth}
\includegraphics[height=2.4in, trim={0.075in 0 0 0},clip]{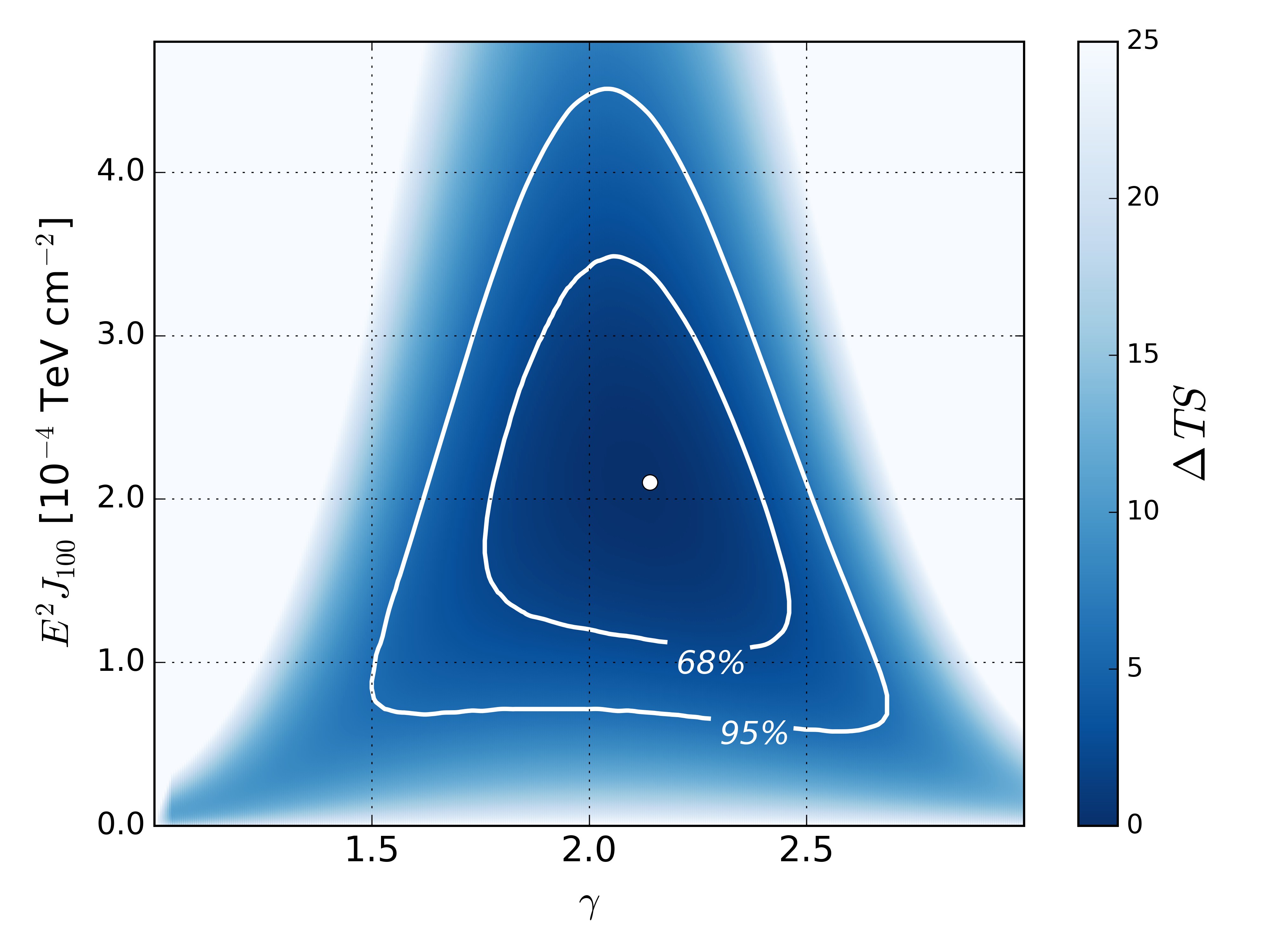}
\caption{}
\label{fig:time_dep_flux_contour}
\end{subfigure}
\hfill
\begin{subfigure}[t]{0.48\textwidth}
\includegraphics[height=2.37in, trim={0in 1.2in 0in 0.47in},clip]{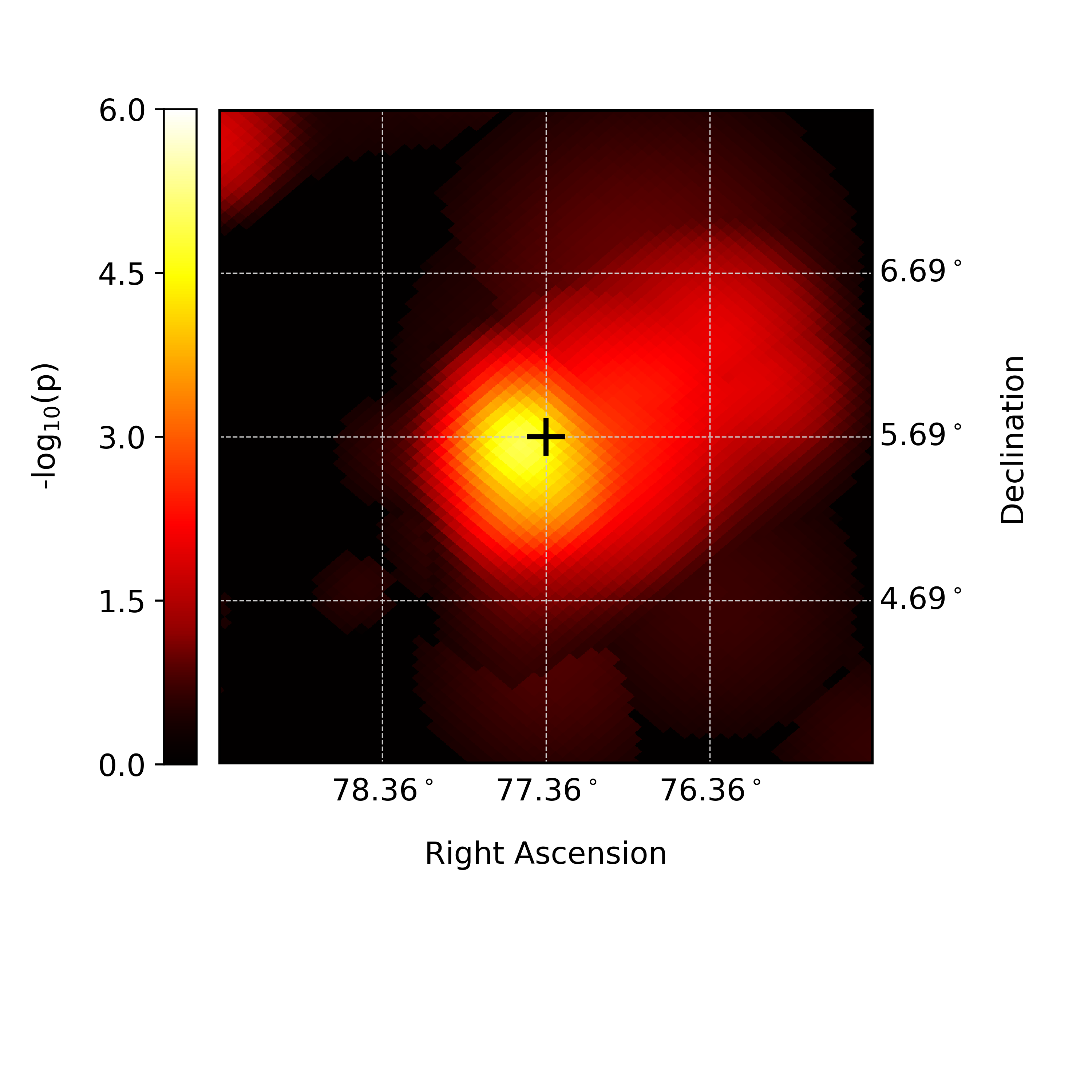}
\caption{}
\label{fig:time_dep_map}
\end{subfigure}
\caption{{\bf Time-dependent analysis results for the IC86b data period (2012-2015)}.
(a) Change in test statistic, $\Delta TS$,  as a function of the spectral index parameter $\gamma$ and the fluence at 100~TeV given by $E^2 J_{100}$. 
The analysis is performed at the coordinates of \TXS/, using the Gaussian-shaped time window and holding the time parameters fixed ($T_0=$~13 December 2014,  $T_\mathrm{W}=110$~days). 
The white dot indicates the best-fitting values.
The contours at 68\% and 95\% confidence level assuming Wilks' theorem \cite{Wilks:1938dza} are shown in order to indicate the statistical uncertainty on the parameter estimates. Systematic uncertainties are not included. (b) Skymap showing the $P$ value of the time-dependent analysis performed at the coordinates of \TXS/ (cross) and at surrounding locations.  The analysis is performed on the IC86b data period, using the Gaussian-shaped time-window.  At each point, the full fit for $(\Phi,\gamma,T_0,T_\mathrm{W})$ is performed.  The $P$ value shown does not include the look-elsewhere effect related to other data periods.  An excess of events is detected consistent with the position of \TXS/.
}
\label{fig:time_dep_result}
\end{center}
\end{figure}

When we estimate the significance of the time-dependent result by performing the analysis at the coordinates of \TXS/ on randomized data sets, we allow in each trial a new fit for all the parameters: 
$\Phi_{100},\gamma,T_0,T_\mathrm{W}$. We find that the fraction of randomized trials that result in a more significant excess than the real data is $7\times 10^{-5}$ for the box-shaped time window and $3\times 10^{-5}$ for the Gaussian time window.
This fraction, once corrected for the ratio of the total observation time to the IC86b observation time (9.5~years / 3~years), results in $P$ values of $2\times 10^{-4}$ and $10^{-4}$, respectively, corresponding to $3.5\sigma$ and $3.7\sigma$.
Because there is no {\it a priori} reason to prefer one of the generic time-windows over the other, we take the more significant one and include a trial factor of 2 for the final significance, which is then $3.5\sigma$.

Outside the 2012-2015 time period, the next most significant excess is found using the Gaussian window in 2017 and includes the \ICA/ event.  This time window is centered at 22 September 2017 with duration $T_\mathrm{W}=19$~days, $\gamma=1.7 \pm 0.6$, and fluence $E^2 J_{100} = 0.2_{-0.2}^{+0.4}\times10^{-4}$ TeV cm$^{-2}$ at~100 TeV. 
No other event besides the \ICA/ event contributes significantly to the best-fit. As a consequence, the uncertainty
on the best-fitting window location and width spans the entire IC86c period, because any window containing \ICA/ yields a similar value of the test statistic.
Following the trial-correction procedure for different observation periods as described above, the significance of this excess is $1.4\sigma$.  If the \ICA/ event is removed, no excess remains during this time period. This agrees with the result of the rapid-response analysis \cite{MeagherICRC17} that is part of the IceCube alert program, which found no other potential astrophysical neutrinos from the same region of the sky during $\pm 7$~days centered on the time of \ICA/.

We performed a time-integrated analysis at the coordinates of \TXS/ using the full 9.5~year-data sample.
The best-fitting parameters for the flux normalization and the spectral index are $\Phi_{100} =$ $(0.8^{+0.5}_{-0.4})\times10^{-16}$ TeV$^{-1}$ cm$^{-2}$ s$^{-1}$ and $\gamma=2.0\pm 0.3$, respectively.
The joint uncertainty on these parameters is shown in Fig.~4a.
The $P$ value, based on repeating the analysis at the same coordinates with randomized data sets, is 0.002\% ($4.1\sigma$), but this is an {\it a posteriori} significance estimate
because it includes the \ICA/ event which motivated performing the analysis at the coordinates of \TXS/. 
An unbiased significance estimate including the event would need to take into account the look-elsewhere effect related to all other possible directions in the sky that could be analyzed.
It is expected that there will be two or three directions somewhere in the northern sky with this significance or greater resulting from the chance alignment of neutrinos \cite{2017ApJ...835..151A}. 
Here we are interested in determining whether there is evidence of time-integrated neutrino emission from \TXS/ besides the \ICA/ event.

If we remove the final data period IC86c, which contains the event, and perform the analysis again using only the first 7 years of data, we find best-fitting parameters that are nearly unchanged: $\Phi_{100} = (0.9^{+0.6}_{-0.5})\times10^{-16}$ TeV$^{-1}$ cm$^{-2}$ s$^{-1}$ and $\gamma=2.1\pm 0.3$, respectively. The joint uncertainty on these parameters is shown in Fig.~4b.
The $P$ value, using only the first 7 years of data, is 1.6\% ($2.1\sigma$), based on repeating the analysis at the same coordinates with randomized data sets.  
These results indicate that the time-integrated fit is dominated by the same excess as found in the time-dependent analysis above, having similar values for the spectral index and 
total fluence ($E^2 J_{100} = 2.0 \times10^{-4}$ TeV cm$^{-2}$ at 100~TeV, over the 7-year period). 
This excess is not significant in the time-integrated analysis because of the additional background during the rest of the 7-year period.

\begin{figure}[ht!]
\begin{center}
\begin{subfigure}[b]{0.49\textwidth}
\includegraphics[width=\textwidth,trim={1cm 0 0 0},clip]{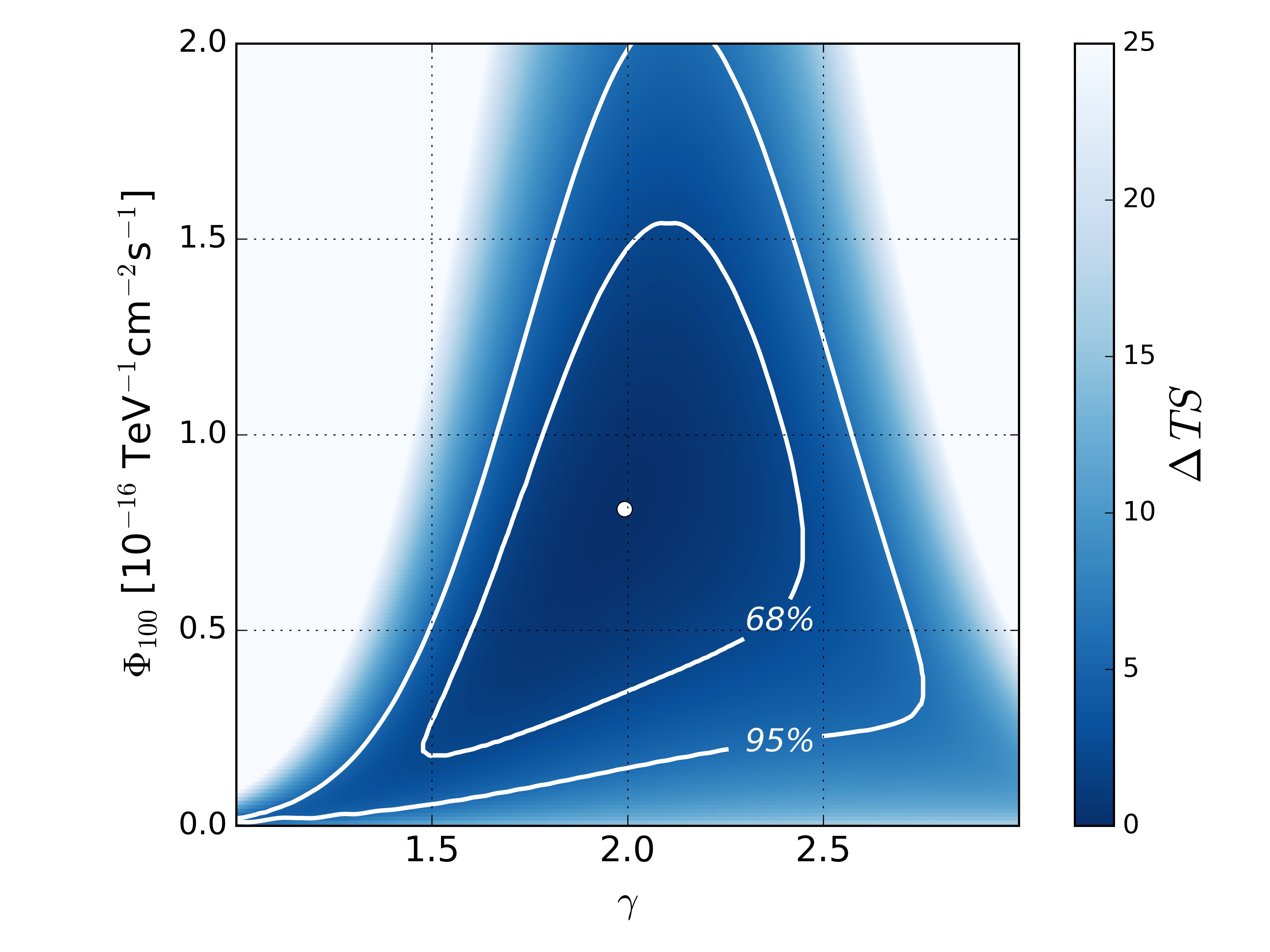}
\caption{}
\label{fig:time_int_flux_contour_9.5}
\end{subfigure}
\begin{subfigure}[b]{0.49\textwidth}
\includegraphics[width=\textwidth,trim={1cm 0 0 0},clip]{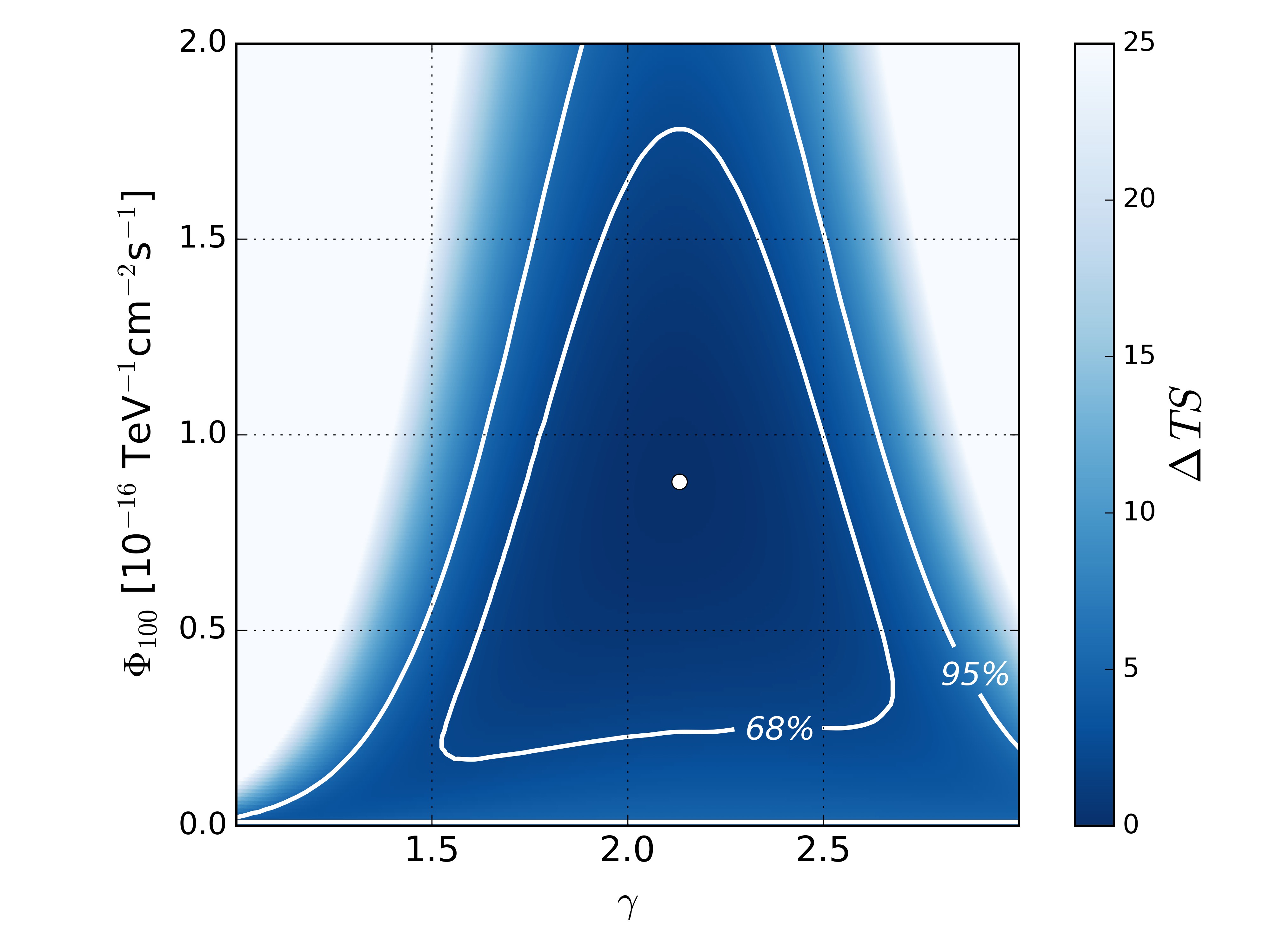}
\caption{}
\label{fig:time_int_flux_contour_7}
\end{subfigure}
\caption{
{\bf Time-integrated analysis results.}  As in 
Fig.~3a,
but for the time-integrated analysis of \TXS/ using (a) the full 9.5 year sample (2008-2017), and (b) the 7 year sample (2008-2015).
}
\label{fig:time_int_results}
\end{center}
\end{figure}

\section*{Blazars as neutrino sources}

The signal identified during the 5-month period in 2014-2015 consists of an estimated $13\pm 5$ muon-neutrino events that are present in addition to the expected background.  The analysis is unbinned, but the mean background at the declination of \TXS/ is useful for comparison purposes; it is 5.8 events in a search bin of radius $1^{\circ}$  during a 158-day time window.  (We use the duration of the box-shaped time-window result for convenience to calculate averages during the flare.) The significance of the excess is due to both the number of events and their energy distribution, with higher-energy events increasing the significance and leading to the best-fitting spectral-index of 2.1, 
in contrast to the lower energy atmospheric neutrino background with spectral index
$\sim 3.7$. At this declination in the sky, the 68\% central energy range in which IceCube is most sensitive to point sources with $E^{-2.1}$ spectra is between 32~TeV and 3.6~PeV.
Assuming that the muon-neutrino fluence ($E^2 J_{100} = (2.1^{+1.0}_{-0.7})\times10^{-4}$ TeV cm$^{-2}$) is one-third of the total neutrino fluence, then the all-flavor neutrino energy fluence is
$(4.2^{+2.0}_{-1.4})\times 10^{-3}$ erg cm$^{-2}$ over this energy range.
With the recent measurement \cite{Paiano:2018qeq} of the redshift of \TXS/ as $z = 0.3365\pm0.0010$, this energy fluence implies that the isotropic neutrino luminosity is 
$(1.2^{+0.6}_{-0.4}) \times 10^{47}$ erg s$^{-1}$ averaged over 158 days. This is higher than the isotropic gamma-ray luminosity during the same period, which is similar to the long-term luminosity between 0.1~GeV and 100~GeV of $0.28\times 10^{47}$ erg s$^{-1}$ averaged over all \FermiLAT/ observations of \TXS/ \cite{MMpaper}. 
Gamma rays are expected to be produced in the same processes that produce neutrinos---for example, when accelerated protons interact with ambient lower-energy photons near the source, producing both neutral pions (which decay to gamma-rays) and charged pions (which decay to neutrinos and leptons). A higher luminosity in neutrinos than in gamma rays
could imply that a substantial fraction of the gamma rays related to the neutrino production are either absorbed or arriving at energies above or below the \FermiLAT/ energy band.

Although \TXS/ is a bright object in gamma rays, it was not previously singled out as a predicted neutrino source.  In the third catalog of active galactic nuclei detected by \FermiLAT/
\cite{Ackermann:2015yfk} 
listing 1773 objects (including those at low galactic latitudes), \TXS/ is among the 50 brightest objects, with an average flux between 1 GeV and 100 GeV of ($6.5\pm0.2) \times 10^{-9}$ photons cm$^{-2}$ s$^{-1}$.
Its measured redshift now makes it one of the most luminous objects known out to the same distance, more than an order of magnitude more luminous than nearby blazars such as Markarian 421, Markarian 501, and 1ES~1959+650.  With respect to these objects, an important observational distinction is the favorable declination of \TXS/.  As the neutrino-nucleon interaction cross-section grows with energy, absorption in Earth becomes considerable for neutrinos above $\sim$100~TeV.  IceCube is most sensitive to high-energy neutrinos from sources at declinations near the equatorial plane, which is viewed along the horizon from the South Pole.
The blazars mentioned above are at more northern declinations, and the likelihood that a neutrino with energy of $\sim$ 300 TeV from one of these 
will be absorbed while traversing Earth is three to five times the likelihood that it will reach the detector.
 The explanation for why \TXS/ is the first blazar associated with a significant neutrino excess may therefore depend on the combination of its intrinsic properties and the observational properties of the detector.

IceCube recently published \cite{2017ApJ...835...45A} 
a search for neutrino emission from the blazars in the second catalog of active galactic nuclei detected by \FermiLAT/ \cite{2011ApJ...743..171A}, 
constraining their contribution to the diffuse astrophysical neutrino flux under different model assumptions.  An upper limit of 27\% was found assuming the diffuse flux that is fit between 10~TeV and 100~TeV with a soft $E^{-2.5}$ spectrum \cite{2015ApJ...809...98A}. 
For an $E^{-2}$ spectrum compatible with the diffuse flux fit above $\sim 200$ TeV \cite{2016ApJ...833....3A}, 
the upper limit is between 40\% and 80\%. The allowed contribution by blazars as a population is larger, because it would include the contribution of fainter and more distant blazars not yet resolved in the catalog. Averaged over 9.5 years, the neutrino flux of \TXS/ by itself corresponds to 1\% of the astrophysical diffuse flux, and is fully compatible with the previous blazar catalog results.

The evidence presented above supports the hypothesis presented in \cite{MMpaper}
that the blazar \TXS/ is a high-energy neutrino source. 
The $3.5\sigma$ evidence for neutrino emission during the 5-month period in 2014-2015
is statistically independent of the evidence presented in \cite{MMpaper}. 
The analysis of the \ICA/ event in \cite{MMpaper} relies on correlation of a single neutrino with electromagnetic activity, whereas the analysis presented here relies only on self-correlation of multiple neutrinos.
The coincidence of an IceCube alert with a flaring blazar, combined with a neutrino flare from the same object in archival IceCube data, pinpoints
a likely source of high-energy cosmic rays.


\subsection*{Supplementary Materials}
www.sciencemag.org\\
Materials and Methods\\
Figs. S1 -- S6\\

\subsection*{Acknowledgments}

{\bf Funding:}
We acknowledge support from the following agencies: USA -- U.S. National Science Foundation-Office of Polar Programs,
U.S. National Science Foundation-Physics Division,
Wisconsin Alumni Research Foundation,
Center for High Throughput Computing (CHTC) at the University of Wisconsin-Madison,
Open Science Grid (OSG),
Extreme Science and Engineering Discovery Environment (XSEDE),
U.S. Department of Energy-National Energy Research Scientific Computing Center,
Particle astrophysics research computing center at the University of Maryland,
Institute for Cyber-Enabled Research at Michigan State University,
and Astroparticle physics computational facility at Marquette University;
Belgium -- Funds for Scientific Research (FRS-FNRS and FWO),
FWO Odysseus and Big Science programmes,
and Belgian Federal Science Policy Office (Belspo);
Germany -- Bundesministerium f\"ur Bildung und Forschung (BMBF),
Deutsche Forschungsgemeinschaft (DFG),
Helmholtz Alliance for Astroparticle Physics (HAP),
Initiative and Networking Fund of the Helmholtz Association,
Deutsches Elektronen Synchrotron (DESY),
and High Performance Computing cluster of the RWTH Aachen;
Sweden -- Swedish Research Council,
Swedish Polar Research Secretariat,
Swedish National Infrastructure for Computing (SNIC),
and Knut and Alice Wallenberg Foundation;
Australia -- Australian Research Council;
Canada -- Natural Sciences and Engineering Research Council of Canada,
Calcul Qu\'ebec, Compute Ontario, Canada Foundation for Innovation, WestGrid, and Compute Canada;
Denmark -- Villum Fonden, Danish National Research Foundation (DNRF);
New Zealand -- Marsden Fund;
Japan -- Japan Society for Promotion of Science (JSPS)
and Institute for Global Prominent Research (IGPR) of Chiba University;
Korea -- National Research Foundation of Korea (NRF);
Switzerland -- Swiss National Science Foundation (SNSF).\\

{\bf Author contributions:}
The IceCube Collaboration designed, constructed and now operates the IceCube Neutrino Observatory. Data processing and calibration, Monte Carlo simulations of the detector and of theoretical models, and data analyses were performed by a large number of collaboration members, who also discussed and approved the scientific results presented here. The manuscript was reviewed by the entire collaboration before publication, and all authors approved the final version.\\

{\bf Competing interests:} There are no competing interests to declare.\\

{\bf Data and materials availability:}
Additional data and resources are available from the IceCube data archive at \url{http://www.icecube.wisc.edu/science/data}. For each data sample these include the events within $3^{\circ}$ of the \TXS/ source coordinates, neutrino effective areas, background rates, and other supporting information in machine-readable formats.\\

\pagebreak
\pagebreak

\begin{center}
\Large{
Supplementary Materials for:\\
Neutrino emission from the direction of the blazar TXS~0506+056 prior to the IceCube-170922A alert
}
\end{center}

\subsection*{IceCube Collaboration$^{\ast}$:}

M.~G.~Aartsen$^{16}$,
M.~Ackermann$^{51}$,
J.~Adams$^{16}$,
J.~A.~Aguilar$^{12}$,
M.~Ahlers$^{20}$,
M.~Ahrens$^{43}$,
I.~Al~Samarai$^{25}$,
D.~Altmann$^{24}$,
K.~Andeen$^{33}$,
T.~Anderson$^{48}$
% ...
\\ 
\\
% ...
$^{12}$ Universit\'e Libre de Bruxelles, Science Faculty CP230, B-1050 Brussels, Belgium \\
% ...
$^{16}$ Dept.~of Physics and Astronomy, University of Canterbury, Private Bag 4800, Christchurch, New Zealand \\
% ...
$^{20}$ Niels Bohr Institute, University of Copenhagen, DK-2100 Copenhagen, Denmark \\
% ...
$^{24}$ Erlangen Centre for Astroparticle Physics, Friedrich-Alexander-Universit\"at Erlangen-N\"urnberg, D-91058 Erlangen, Germany \\
$^{25}$ D\'epartement de physique nucl\'eaire et corpusculaire, Universit\'e de Gen\`eve, CH-1211 Gen\`eve, Switzerland \\
% ...
$^{33}$ Department of Physics, Marquette University, Milwaukee, WI, 53201, USA \\
% ...
$^{43}$ Oskar Klein Centre and Dept.~of Physics, Stockholm University, SE-10691 Stockholm, Sweden \\
% ...
$^{48}$ Dept.~of Physics, Pennsylvania State University, University Park, PA 16802, USA \\
% ...
$^{51}$ DESY, D-15738 Zeuthen, Germany \\
% ...
$^\ast$E-mail: analysis@icecube.wisc.edu

\section*{Materials and Methods}

\renewcommand\thetable{S\arabic{table}}
\renewcommand\thefigure{S\arabic{figure}}
\renewcommand\theequation{S\arabic{equation}}

\setcounter{table}{0}
\setcounter{figure}{0}
\setcounter{equation}{0}

\textbf{Time-integrated analysis method}

The time-integrated analysis uses the unbinned maximum likelihood technique
described in \cite{2008APh....29..299B} to quantify spatial clustering of neutrino events on the sky. Energy information is also used in the analysis, primarily as a means to help separate signal and background rather than to characterize the energy distribution of the neutrino signal itself.
In this method the unbinned likelihood is defined as a product over all neutrino events in the data sample:
\begin{equation}
\mathcal{L}(\Phi_{100}, \gamma) = \prod_i \Big( \frac{n_S(\Phi_{100}, \gamma)}{N} \mathcal{S}(\mathbf{x}_S, \mathbf{x}_i, \sigma_i, E_i; \gamma) + \Big(1 - \frac{n_S(\Phi_{100}, \gamma)}{N}\Big) \mathcal{B}(\mathrm{sin} \, \delta_i, E_i) \Big)
\end{equation}
\noindent where $\mathcal{S}$ represents the probability distribution function (PDF) of signal events from a point source, $\mathcal{B}$ represents the PDF of background events, $N$ is the total number of neutrino events, and $n_S(\Phi_{100}, \gamma)$ is the corresponding number of signal neutrinos in the sample for a signal flux model of the form $\Phi(E) = \Phi_{100} (E / 100 \, \mathrm{TeV})^{-\gamma}$ where $E$ is the true neutrino energy, $\gamma$ is the spectral index, and $\Phi_{100}$ is the flux normalization at 100 TeV.

$\mathcal{S}$ depends on the position of the source $\mathbf{x}_S$, given as a vector in right ascension and declination, the reconstructed direction of the neutrino event $\mathbf{x}_i$, and the spectral index $\gamma$ from the power-law flux model. It can be factorized into a spatial component and an energy component. The spatial component is described by a two-dimensional Gaussian, exp$(- \lvert \mathbf{x}_S - \mathbf{x}_i \rvert^2 / 2 \, \sigma_i^2 ) \, / \, 2\pi\sigma_i^2$, where $\sigma_i$ is the directional reconstruction uncertainty for the $i^{\mathrm{th}}$ neutrino event. The energy component $\mathcal{E}_S(E_i, \mathrm{sin} \, \delta_i; \gamma)$ accounts for the probability of observing a neutrino event with reconstructed energy $E_i$ (muon energy proxy, defined below), at a reconstructed declination of $\delta_i$ for the signal flux model with spectral index $\gamma$. $\mathcal{E}_S$ is a function of declination because the effective area of IceCube is declination dependent.The full form is:
\begin{equation}
\mathcal{S}(\mathbf{x}_S, \mathbf{x}_i, \sigma_i, E_i; \gamma) =
     \frac{1}{2\pi\sigma_i^2} \, \mathrm{e}^{ - \frac{\lvert \mathbf{x}_S - \mathbf{x}_i \rvert^2}{2\sigma_i^2} } 
     \times \mathcal{E}_S(E_i, \mathrm{sin} \, \delta_i; \gamma).
\end{equation}
\noindent 

We construct $\mathcal{B}$ from individual spatial and energy components for background events as well. The spatial term is estimated using experimental data and depends only on the event's declination $\delta_i$, because the probability distribution of background events is uniform in right ascension. The spatial PDF for background, $\mathcal{P_B}(\mathrm{sin} \, \delta_i)$, is equal to the event density per solid angle, divided by the total number of events in the sample. 
The background energy PDF $\mathcal{E}_B(E_i, \mathrm{sin} \, \delta_i)$ is also estimated directly from experimental data by measuring the fraction of events with energy proxy $E_i$ at a reconstructed declination of $\mathrm{sin} \, \delta_i$. The full form of $\mathcal{B}$ is:
\begin{equation}
\mathcal{B}(\mathrm{sin} \, \delta_i, E_i) = \mathcal{P}_\mathcal{B}(\mathrm{sin} \, \delta_i) 
     \times \mathcal{E}_B(E_i, \mathrm{sin} \, \delta_i).
\end{equation}

The ratio $\mathcal{S}/\mathcal{B}$ evaluated for a single event is referred to as the Event Weight.  The main contribution to the likelihood and the final significance comes from events with large values of $\mathcal{S}/\mathcal{B}$ (see e.g.\ %
Fig.~2).
For a given $\Phi(E)$, the number of signal events $n_S$ expected in the data depends upon the exposure time $T$ and the declination-dependent effective area, $A_{eff}(E, \mathrm{sin} \, \delta_i)$, that characterizes the detector and data sample. It is given by:
\begin{equation}
\label{Eq:ns_from_Aeff}
n_S = T \times \int_{100 \, \mathrm{GeV}}^{1 \, \mathrm{EeV}} A_{eff}(E, \mathrm{sin} \, \delta_i) \, \Phi(E)  dE
\end{equation}
where the integration is over the range of true neutrino energies relevant for the data samples.
We calculate $n_S(\Phi_{100}, \gamma)$ from Eq.~\ref{Eq:ns_from_Aeff} for the case where $\Phi(E) = \Phi_{100} (E / 100 \, \mathrm{TeV})^{-\gamma}$.

Properties such as effective area and exposure time vary for each of the data samples shown in Table 1, so we compute $\mathcal{L}$ separately for each, and combine them as a product:
\begin{equation}
\mathcal{L}(\Phi_{100}, \gamma) = \prod_j \mathcal{L}_j(\Phi_{100}, \gamma)
\end{equation}
where $j$ denotes the data sample. The case $\mathcal{L}(\Phi_{100} = 0)$ represents the null hypothesis of no point source signal present (in which case $\gamma$ is not defined). We form the test statistic $TS = 2 \, \mathrm{log}( \mathcal{L}(\Phi_{100}, \gamma) / \mathcal{L}(\Phi_{100} = 0) )$ as a likelihood ratio relative to the null hypothesis and maximize it as a function of its two parameters over the ranges $\Phi_{100} \ge 0$, $\gamma \in [1, 4]$ to find the best-fitting values of $\Phi_{100}$ and $\gamma$.

We estimate the probability distribution of the test statistic under the null hypothesis by
performing the time-integrated analysis at the location of \TXS/
on samples of randomized data. Each sample is produced by assigning random
values of right ascension, from 0$^\circ$ to 360$^\circ$, to events in the original data set with a random number generator.
The significance of the result for the real data is given by the p-value, defined as the fraction of randomized trials with $TS \ge TS_\mathrm{data}$.  The test-statistic distribution for the time-integrated analysis performed on 5 million randomized data sets is shown in Fig.~\ref{fig:time_int_trials}.
The advantage of using randomized experimental data samples to simulate the null hypothesis is that the resulting significance estimate is generally conservative and robust against systematic errors due for example to imperfect detector simulation of the background. \\

\noindent \textbf{Time-dependent analysis method}

The time-dependent analysis uses the unbinned maximum-likelihood technique described in
\cite{timeDepMethod2010APh....33..175B} to
search for a point source signal in which the neutrinos also cluster in time. In the model-independent version of this analysis, minimal assumptions about the time structure of the neutrino signal are made. It is only assumed that the emission is clustered around some time $T_0$ with some duration $T_\mathrm{W}$, where these are parameter values to be determined by the best fit to the data. As with the use of energy information in the analysis, the search for time-clustering is primarily a way to improve the chance of identifying a possibly time-dependent neutrino signal, rather than to provide a detailed characterization of time variability.

We consider two functional forms for the temporal distribution of events, a box function and a one-dimensional Gaussian function.
These both provide reasonable fits to generic transient signals in the low statistics regime
where the small number of recorded signal events provide minimal information about the exact shape. The box function is given by
\begin{equation}
\mathcal{T}_S(t_i; T_0, T_\mathrm{W}) = \frac{1}{T_\mathrm{W}}  \quad (T_0 - T_\mathrm{W} / 2 < t_i < T_0 + T_\mathrm{W} / 2) \quad \mathrm{BOX}
\end{equation}
where $T_0$ is the central time and $T_\mathrm{W}$ is the full width of the box shape. The Gaussian
shape is described by the same parameters but with the following functional form
\begin{equation}
\mathcal{T}_S(t_i; T_0, T_\mathrm{W}) = \frac{1}{\sqrt{2 \pi (T_\mathrm{W} / 2)^2}} \, e^{- \frac{(t_i - T_0)^2}{2 (T_\mathrm{W} / 2)^2} } \quad \mathrm{GAUSSIAN}
\end{equation}

The PDF for the time distribution of background events $\mathcal{T}_B$ is given to good approximation by $1 / T$, where $T$ is the total observation time of the data sample, 
because IceCube maintains essentially uniform efficiency with time during each individual observation period \cite{2017JInst..12P3012A}.

The unbinned likelihood for this method is similar to the time-integrated likelihood with the functions $\mathcal{S}$ and $\mathcal{B}$ defined as before, and now including signal and background temporal terms. It is defined as a product over all neutrino events in the data sample
\begin{equation}
\mathcal{L}(\Phi_{100}, \gamma, T_0, T_\mathrm{W}) = \prod_i \Big( \frac{n_S(\Phi_{100}, \gamma)}{N} \mathcal{S}
\times \mathcal{T}_S(t_i; T_0, T_\mathrm{W})
+ \Big(1 - \frac{n_S(\Phi_{100}, \gamma)}{N}\Big) \mathcal{B}
\times \mathcal{T}_B
 \Big).
\end{equation}

We compute the time-dependent likelihood separately for each of the data periods shown in Table 1
under the assumption that we are looking for neutrino emission on timescales less than the
$\sim$~year durations of the data sets. This is because significant emission on longer
timescales will appear in the time-integrated analysis. 
As before, the test statistic is constructed using a likelihood ratio of the signal and null hypotheses:
\begin{equation}
TS = 2 \, \mathrm{log} \Big[ \, \frac{T_\mathrm{W}}{\mathrm{T}} \times \frac{\mathcal{L}(\Phi_{100}, \gamma, T_0, T_\mathrm{W})}{\mathcal{L}(\Phi_{100} = 0)} \, \Big]
\end{equation}
where the additional factor $T_\mathrm{W} / \mathrm{T}$ corrects
for the look-elsewhere effect due to choosing a time window of width $T_\mathrm{W}$ from within
a sample with observation time $T$.
As explained in  \cite{timeDepMethod2010APh....33..175B} where it is derived, the form of this correction arises from the fact that there are more possible short time windows than long ones, with correspondingly more possibilities for a cluster to occur in a short window by chance.
We maximize the test statistic as a function of its four parameters in order to find
the best-fit values of $\Phi_{100}$, $\gamma$, $T_\mathrm{W}$, and $T_0$.
This is done over the ranges $\Phi_{100} \ge 0$, $\gamma \in [1, 4]$,  $T_\mathrm{W} < T / 2$, $T_0 \in [T_\mathrm{min}, T_\mathrm{max}]$
where $T_\mathrm{min}$ and $T_\mathrm{max}$ are the start and end times of the data sample.

Again, we estimate the probability distribution of the test statistic under the null hypothesis by performing the time-dependent analysis
at the location of \TXS/ on samples of randomized data. In the time-dependent case, random samples 
are produced by randomly re-assigning the times among the events
within the same data period used to compute the likelihood. The equatorial coordinates 
are then recomputed using the new times.  All other event properties remain unchanged.
Fig.~\ref{fig:time_dep_trials} shows the resulting test-statistic distributions for the analysis applied to randomized IC86b data samples using both the box and the Gaussian window shapes.

Since we perform the likelihood analysis separately on six data samples, there is an additional look-elsewhere correction when we choose the sample with the best p-value. For small p-values (i.e.~if the result is significant) this correction can be made by multiplying the p-value by $T_\mathrm{total} / T$, where $T$ is the observation time of the chosen sample, and $T_\mathrm{total}$ is the combined observation time of all of the samples. This is an extension of the same look-elsewhere factor described above, from \cite{timeDepMethod2010APh....33..175B}. \\

\noindent \textbf{Muon energy and neutrino energy}

While traversing the ice, muons above $\sim 1$ TeV experience stochastic energy losses due to pair production, bremsstrahlung, and photo-nuclear interactions. The magnitude of these energy losses is inferred from the pattern and intensity of Cherenkov photons recorded by the optical sensors, allowing an estimate of the muon's energy. Inferring the energy of the original neutrino from the muon energy introduces significant uncertainties, because the muon can travel kilometers in the ice and only part of the track is observed as it passes through the instrumented detector volume. In most cases, the muon is created in a neutrino interaction that occurred at an unknown distance before the muon arrived at the detector region, so the observed muon energy that remains is in general a lower bound on the neutrino energy.

The likelihood analysis does not attempt to estimate the original neutrino energy for each event. Instead, for the signal energy PDF, it relies on the distribution of expected muon energies which is determined by detector simulation using the neutrino flux model specified in the signal hypothesis.  The background energy PDF is determined by the muon energy distribution of the experimental data itself.  This allows the likelihood ratio test to be performed directly using the reconstructed muon-energy observables.

For a single event, it is possible to use the reconstructed muon energy to estimate the original neutrino energy, with large uncertainties. Fig.~\ref{fig:Energy_TrueNu_to_RecoMu} shows an example of the distribution of reconstructed muon energies for a range of different neutrino energies based on detector simulations of the IC86b sample for neutrinos arriving from the declination of \TXS/. Given the reconstructed muon energy of a single event, the probability distribution for the original neutrino energy depends on the assumed prior distribution of neutrino energies. In Fig.~\ref{fig:Energy_RecoMu_to_TrueNu_E-2_1} this is shown for the same simulation and assuming that the neutrino flux follows a power-law spectrum $dN/dE \propto E^{-2.1}$.  A specific example for the case of a reconstructed muon energy of 10 TeV in the IC86b sample, at the declination of \TXS/, is shown in Fig.~\ref{fig:Energy_TrueNu_for_Reco_10TeV} for different assumed power-law spectra.

The muon energy reconstruction algorithms are different across the six data samples, as are the expected distributions of muon energies due to different event selections and detector configurations.  The muon energy observable is generically referred to as the Muon Energy Proxy, but it is strictly only comparable with other proxy values from the same data sample.  Details of the event selections and reconstruction algorithms for each sample are described in \cite{Aartsen:2016lmt} 
and references therein.

\newpage

\begin{figure}[ht!]
\begin{center}
\includegraphics[width=0.7\textwidth]{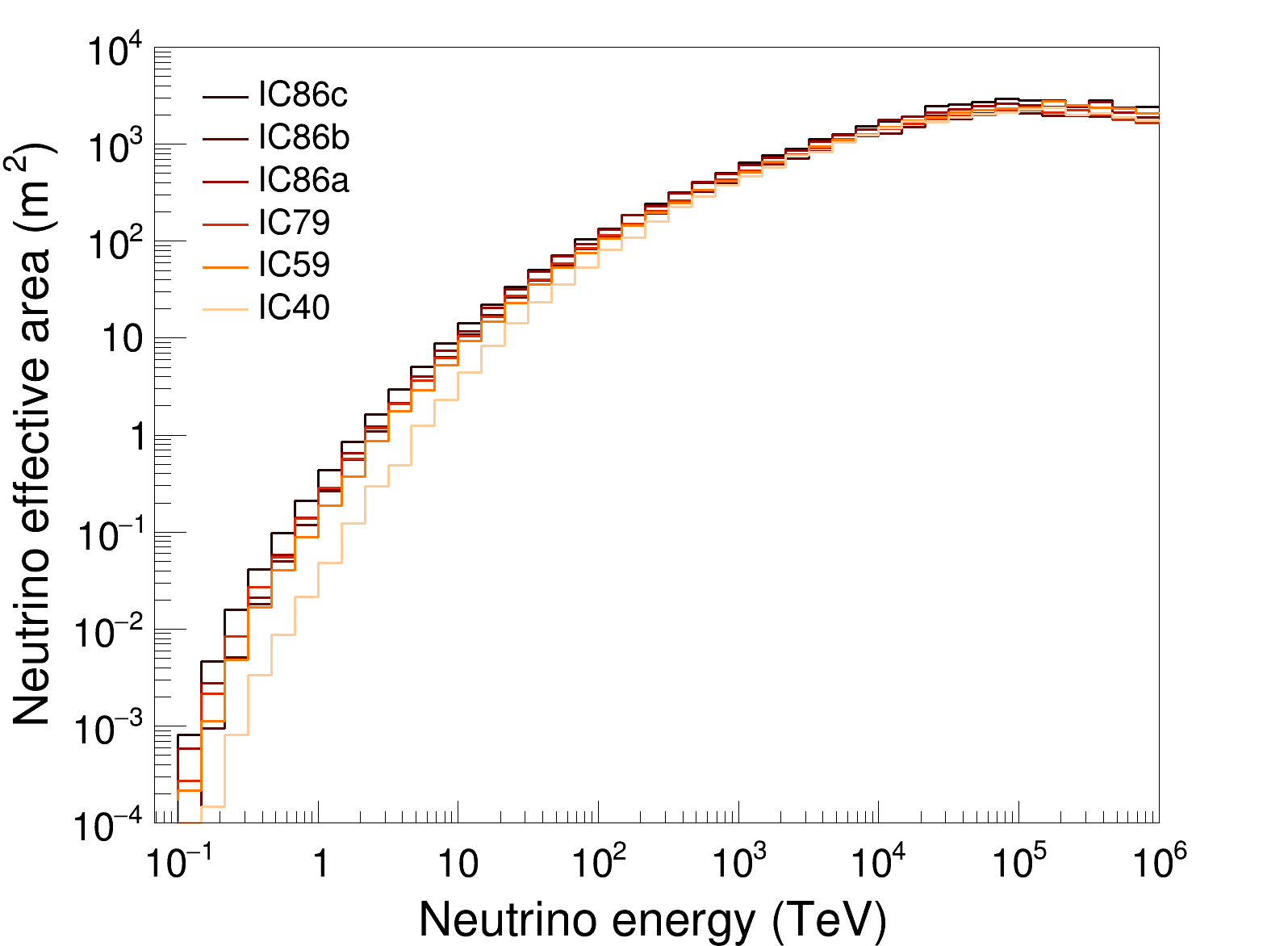}
\caption{{\bf Effective areas for each data sample}.
Each curve represents the effective area of IceCube to muon neutrino events
at the declination of \TXS/ for the different detector configurations.}
\label{fig:eff_areas}
\end{center}
\end{figure}

\newpage

\begin{figure}[ht!]
\begin{center}
\includegraphics[width=0.5\textwidth]{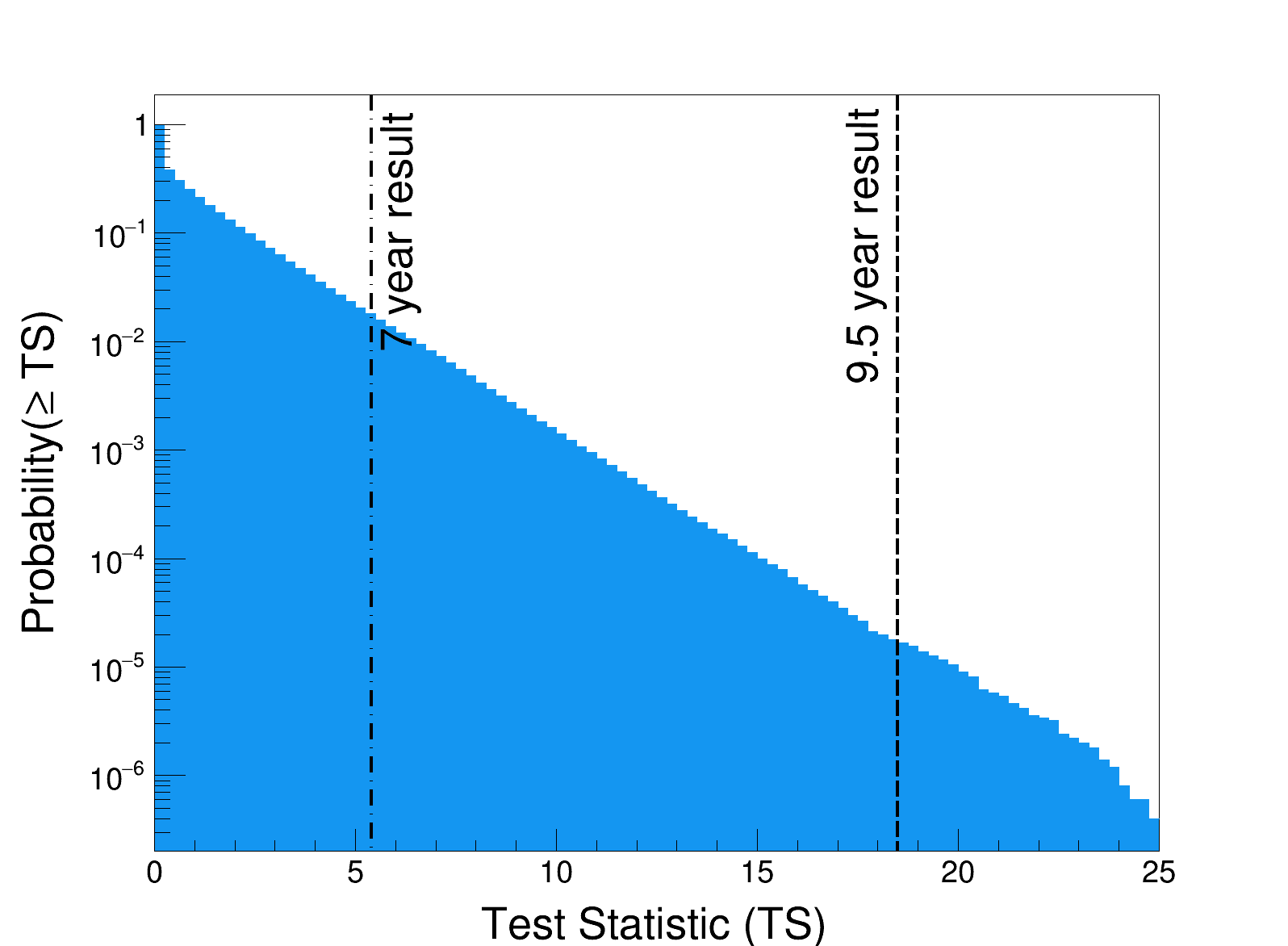}
\caption{{\bf Time-integrated analysis test statistic distribution.} 
The blue curve corresponds to the probability of observing at least the test statistic value $TS$
from the time-integrated analysis performed on randomized data.
The dashed line marks the $TS$ value observed at the location of \TXS/ with 9.5 years of data, including the \ICA/ event.
The dash-dotted line marks the $TS$ value observed at the location of \TXS/ with 7 years of data (IC86c removed), which
does not contain the \ICA/ event.}
\label{fig:time_int_trials}
\end{center}
\end{figure}

\begin{figure}[ht!]
\begin{center}
\begin{subfigure}[t]{0.48\textwidth}
\includegraphics[width=\textwidth]{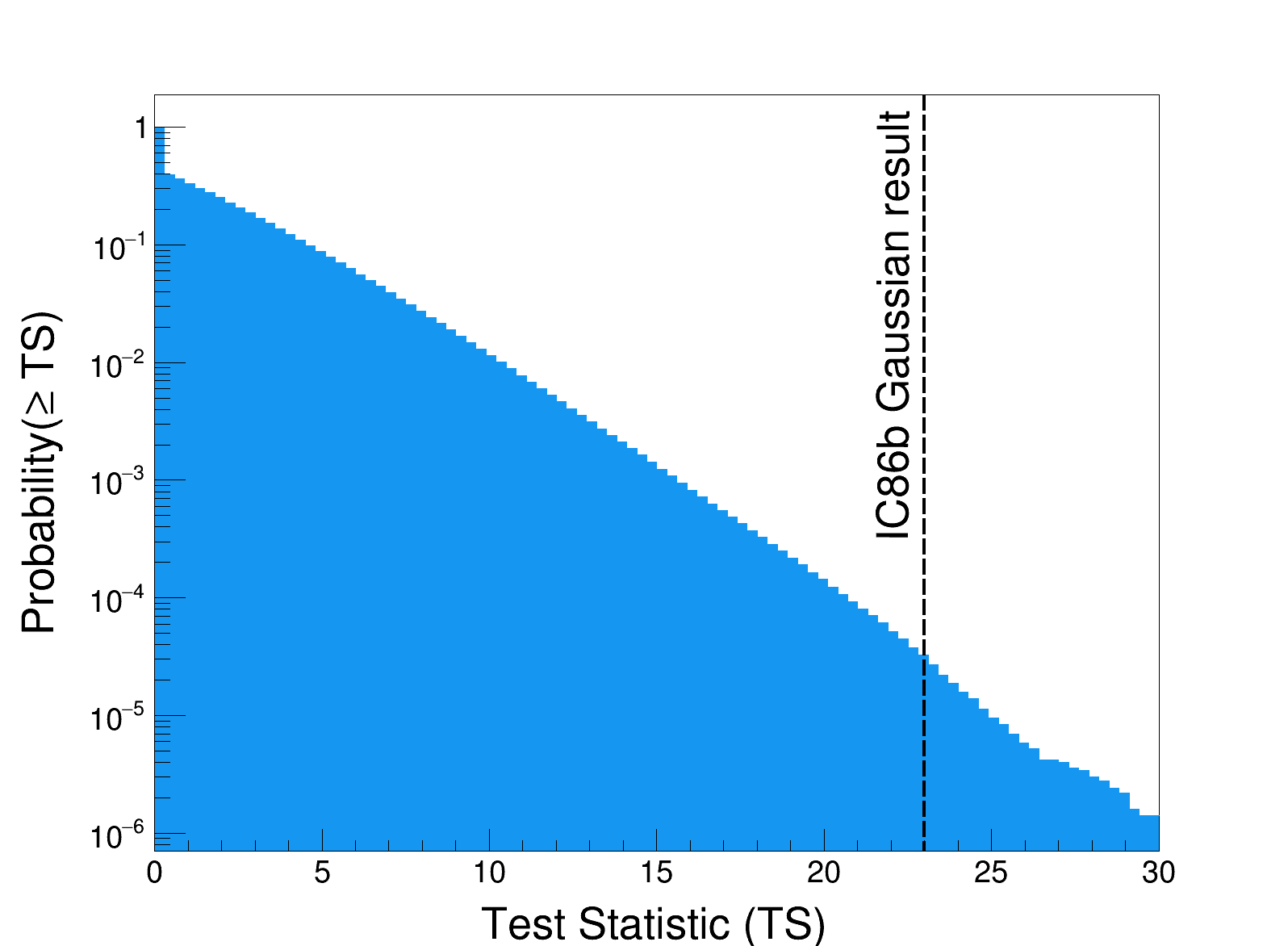}
\caption{}
\end{subfigure}
\begin{subfigure}[t]{0.48\textwidth}
\includegraphics[width=\textwidth]{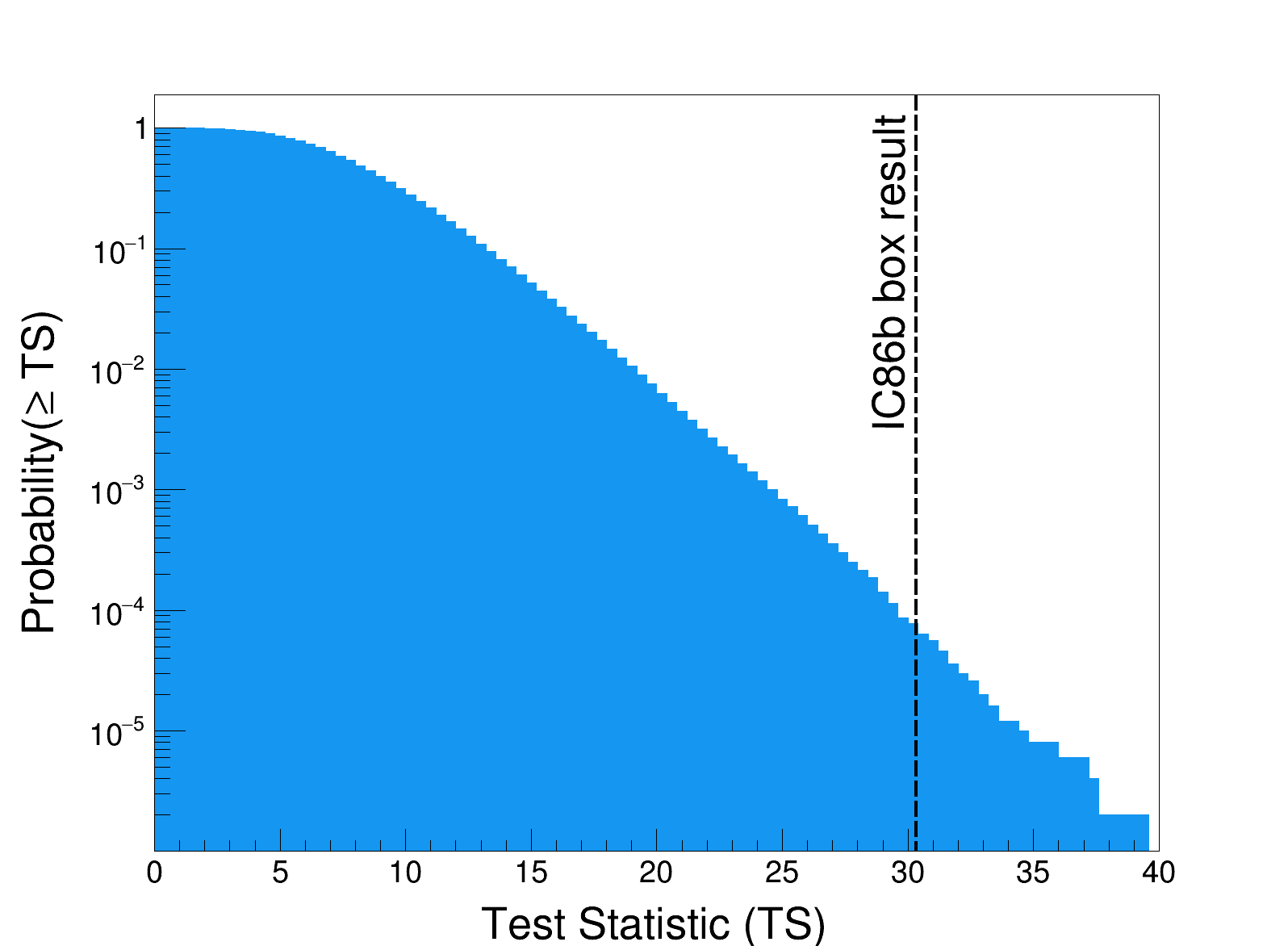}
\caption{}
\end{subfigure}
\caption{{\bf Time-dependent analysis test statistic distributions.} 
The blue curve corresponds to the probability of observing at least the test statistic value $TS$
from the time-dependent analysis performed on randomized data with (a) the Gaussian window and (b) the box window.
The dashed lines mark the $TS$ values observed at the location of \TXS/ with the Gaussian and box windows during the IC86b period.}
\label{fig:time_dep_trials}
\end{center}
\end{figure}

\newpage

\begin{figure}[ht!]
\begin{center}
\includegraphics[width=0.5\textwidth]{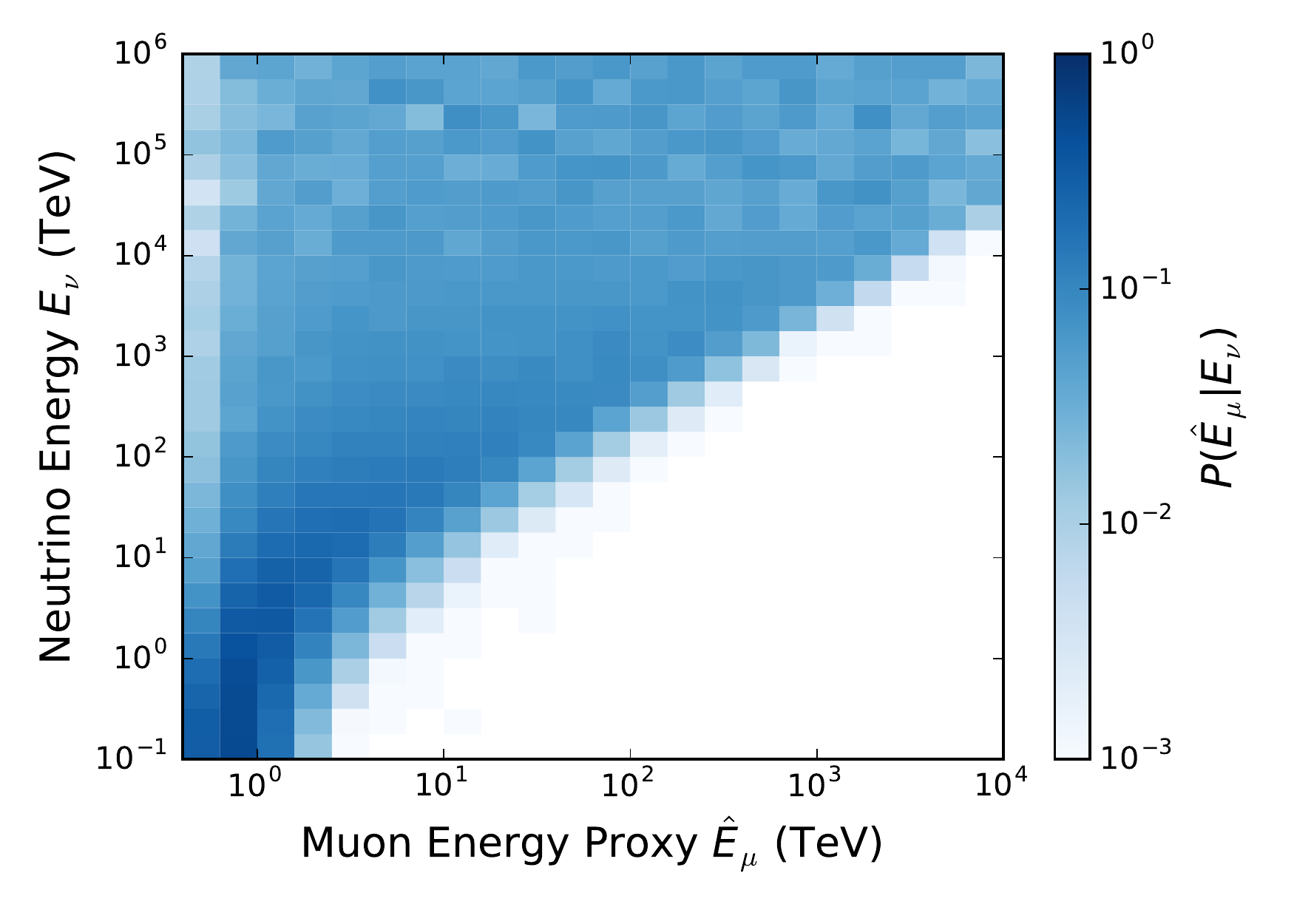}
\caption{{\bf Distribution of Muon Energy Proxy for true neutrino energies.} Distribution of reconstructed muon energies (Muon Energy Proxy) for different neutrino energies specified on the y-axis. Each horizontal slice in neutrino energy $E_\nu$ has been individually normalized. The neutrinos were simulated from the declination of \TXS/ with the same event selection and reconstruction algorithms as applied to the real data for the IC86b sample. 
}
\label{fig:Energy_TrueNu_to_RecoMu}
\end{center}
\end{figure}

\begin{figure}[ht!]
\begin{center}
\includegraphics[width=0.5\textwidth]{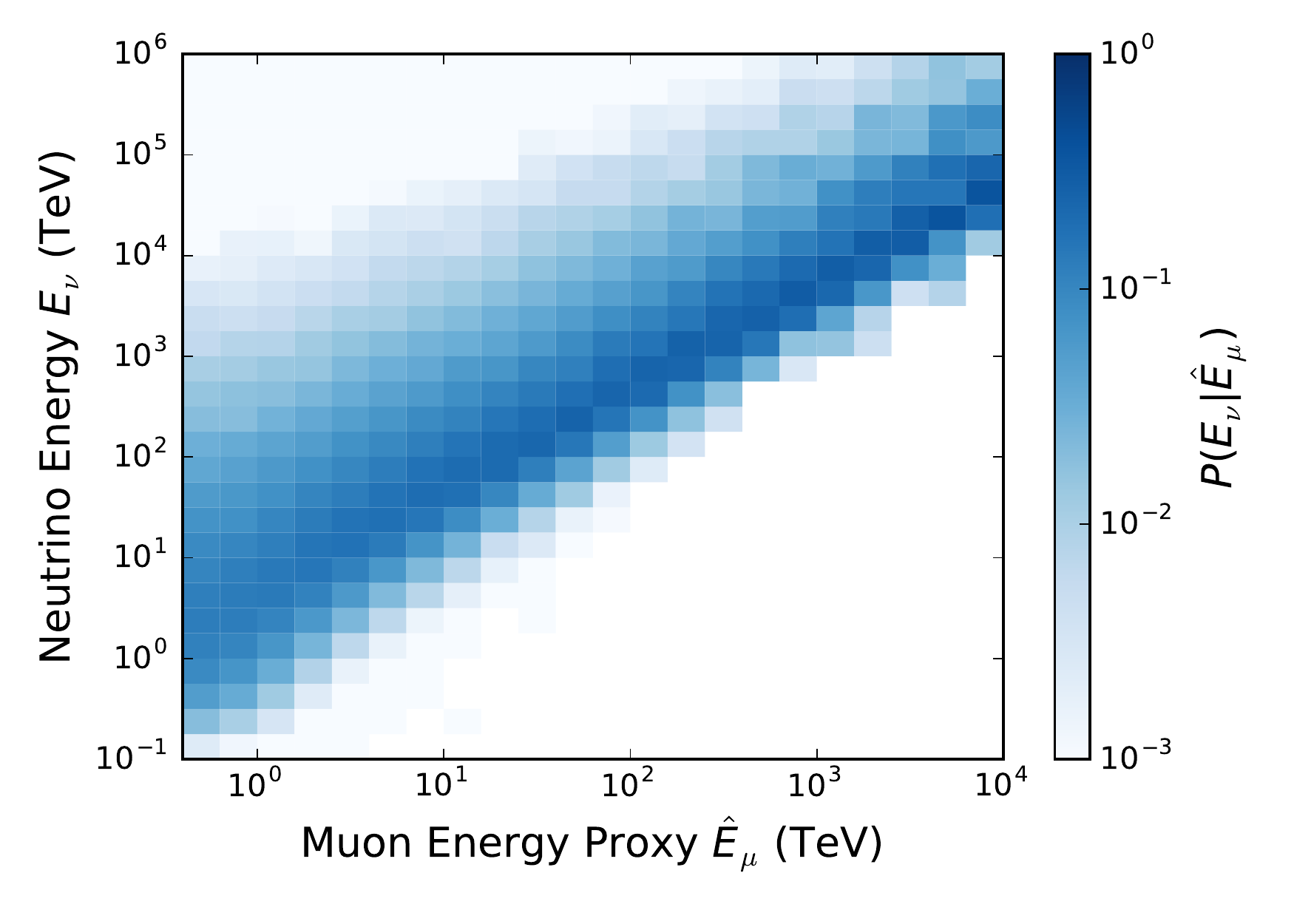}
\caption{{\bf Distribution of true neutrino energy as a function of Muon Energy Proxy.} Distribution of true neutrino energies for different observed values of the Muon Energy Proxy specified on the x-axis. The same simulation was used for IC86b as in Fig.~\ref{fig:Energy_TrueNu_to_RecoMu}, with the neutrino energy distribution weighted according to a power-law spectrum $dN/dE \propto E^{-2.1}$. Each vertical slice in Muon Energy Proxy was then individually normalized. Assuming this prior on the neutrino energy distribution, an observed value of the Muon Energy Proxy determines the probability distribution of the true neutrino energy.
}
\label{fig:Energy_RecoMu_to_TrueNu_E-2_1}
\end{center}
\end{figure}

\newpage

\begin{figure}[ht!]
\begin{center}
\includegraphics[width=0.5\textwidth]{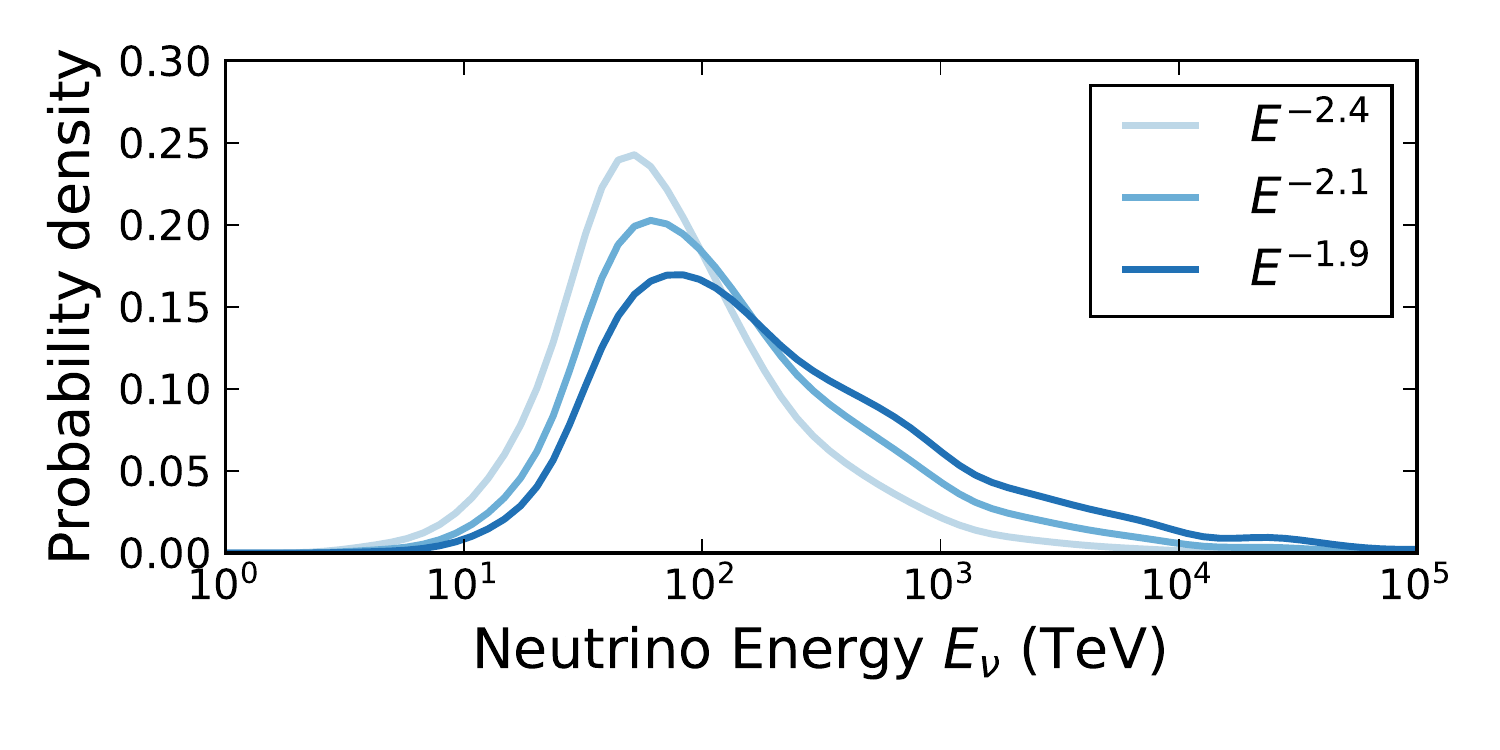}
\caption{{\bf Estimate of neutrino energy for Muon Energy Proxy of 10 TeV.} Probability distribution of neutrino energies corresponding to a Muon Energy Proxy value of 10~TeV, under different assumptions of the neutrino spectral energy distribution. The same simulation was used as in Fig.~\ref{fig:Energy_TrueNu_to_RecoMu}.
}
\label{fig:Energy_TrueNu_for_Reco_10TeV}
\end{center}
\end{figure}

\end{document}